\documentstyle[epsbox]{article}

\def\abstract#1{\vskip 7mm 
        \begin{center}{\large Abstract}\par \smallskip
                \begin{minipage}[c]{12cm}
                        \small #1
                \end{minipage}
        \end{center}
}
\def\title#1{\begin{center}{\Large\bf #1}\end{center}}
\def\author#1{\vskip 5mm \begin{center}{#1}\end{center}}
\def\address#1{\begin{center}{\it #1}\end{center}}
\makeatletter
\@ifundefined{lesssim}{\def\lesssim{\mathrel{\mathpalette\vereq<}}}{}
\@ifundefined{gtrsim}{\def\gtrsim{\mathrel{\mathpalette\vereq>}}}{}
\def\vereq#1#2{\lower3pt\vbox{\baselineskip1.5pt \lineskip1.5pt
\ialign{$\m@th#1\hfill##\hfil$\crcr#2\crcr\sim\crcr}}}
\makeatother

\newcommand{\mpl}{m_{\rm Pl}}

\newcommand{\xx}{1-6\xi}

\begin{document}

\def\thefootnote{\fnsymbol{footnote}}
\title{%
Topological Defect Inflation\footnote
{This talk is based on Refs.\cite{SNH,SYM,SY}.}
}
\author{%
Nobuyuki Sakai\footnote{E-mail:sakai@yukawa.kyoto-u.ac.jp}
}
\address{%
Yukawa Institute for Theoretical Physics, Kyoto University,
Kyoto 606-8502
}
\abstract{
We address some issues of topological defect inflation.
(1) We clarify the causal structure of an inflating magnetic monopole.
The spacetime diagram shows explicitly that this model is free from the
``graceful
exit'' problem, while the monopole itself undergoes ``eternal inflation''.
(2) We extend the study of inflating topological defects to 
Brans-Dicke gravity. Contrary to the case of Einstein gravity, any 
inflating monopole eventually shrinks and takes a stable configuration.
(3) We reanalyze chaotic inflation with a non-minimally coupled massive scalar 
field. We find a new solution of domain wall inflation, which relaxes
constraints 
on the coupling constant for successful inflation.
}

\section{Introduction}

Guendelman \& Rabinowitz, Linde, and Vilenkin independently claimed
that topological defects expand exponentially if the vacuum 
expectation value of the Higgs field $\eta$ is of the order of the Planck 
mass $\mpl$ \cite{GR,LV}. Their arguments
were later verified in numerical simulations \cite{SSTM,Sak}: it was
shown in particular that the critical value of $\eta$
for domain walls and global monopoles is $\eta_{\rm inf}\approx0.33\mpl$,
regardless
of initial conditions.

Because this ``topological defect inflation'' (TDI) model is free
from the fine-tuning problem of initial conditions, it has been attracting
attention. In particular, recently it has been argued that TDI
takes place in some of the plausible models in particle physics \cite{EI}.

In this paper, concerning the TDI model, we address the following three
issues.\\
(1) Causal structure of an inflating magnetic monopole \cite{SNH}.\\
(2) TDI in Brans-Dicke (BD) theory \cite{SYM}.\\
(3) TDI induced by non-minimally coupled massive scalar field \cite{SY}.

\section{Causal Structure of an Inflating Magnetic Monopole}

Among various topological defects, spacetime solutions of magnetic monopoles 
have been studied the most intensively in the literature \cite{VG,OLB,TMT}.
This originated from the rather mathematical interest in static
solutions with non-Abelian hair \cite{VG}. It was shown \cite{OLB} that
static regular solutions are nonexistent if $\eta$ is larger than a critical
value
$\eta_{\rm sta}(\sim\mpl)$. Our previous work \cite{Sak} revealed the
properties 
of monopoles for $\eta>\eta_{\rm sta}$. There are three types of 
solutions, depending on coupling constants and initial configuration: 
a monopole either expands, collapses into a black hole, or comes to 
take a stable configuration. The first type corresponds to TDI.

Recently, the causal structure of an inflating magnetic monopole was
discussed in Ref.\cite{BTV}.
The spacetime diagrams in Ref.\cite{BTV} showed, for instance, that
the inflationary boundary expands along outgoing null geodesics,
that is, no observer can exit from an inflationary region.
The above argument would be fatal to TDI because it
implies that reheating never occurs. Therefore, the spacetime
structure of TDI deserves close examination.
Here we discuss the causal structure of an inflating magnetic 
monopole, based on the numerical solution in Ref.\cite{Sak}.

In order to see the spacetime structure for numerical solutions, we observe
the signs of the expansion of a null geodesic congruence. For a spherically
symmetric metric,
\begin{equation}\label{metric}
ds^2=-dt^2+A^2(t,r)dr^2+B^2(t,r)r^2(d\theta^2+\sin^2\theta d\varphi^2),
\end{equation}
an outgoing $(+)$ or ingoing $(-)$ null vector is given by
$k^{\mu}_{\pm}=(1,\pm A^{-1},0,0)$,
and its expansion $\Theta_{\pm}$ is written as
\begin{equation}\label{expansion}
\Theta_{\pm} =
{k^{\theta}_{\pm;\theta}}+{k^{\varphi}_{\pm;\varphi}}
={2\over B}\bigg({\partial B\over\partial t}
\pm{1\over Ar}{\partial (Br)\over\partial r}\biggl),
\end{equation}
which is defined as the trace of a {\it projection} of $k^{\mu}_{;\nu}$
onto a relevant 2-dimensional surface \cite{NOK}.
We define an ``apparent horizon'' as the surface with $\Theta^+=0$ or
$\Theta^-=0$
\footnote{
In the literature an ``apparent horizon'' usually refers to the outermost
surface with $\Theta^{+}=0$ in an asymptotically flat spacetime.
In this article, however, we call any marginal surface with
$\Theta^+=0$ or $\Theta^-=0$ an apparent horizon.}.
We label those surfaces as S1, S2, \ etc. in our figures.

In Fig.\ 1(a), we plot the trajectories of monopole boundaries and of apparent
horizons in terms of the proper distance from the center: $X\equiv\int^r_0
Adr'$.
Here we define the boundary in two ways: $X_{\phi}$ as the
position of $\phi=\eta/2$ and $X_w$ as the position of $w=1/2$, where
$\phi$ and $w$ are the Higgs field and a gauge-field function, respectively.
An apparent horizon S1 almost agrees with $X=H_c^{-1}\equiv[8\pi
V(0)/3\mpl^2]^{-1/2}$,
which implies that the monopole core is almost de Sitter spacetime.
Figure 1(a) also illustrates that a wormhole structure with black hole horizons
appears
around an inflating core. Because the inflating core becomes causally
disconnected from the outer universe, such an isolated region is
called a ``child universe''. The production of child universes was
originally discussed by Sato, Sasaki, Kodama, and Maeda \cite{SSKM}
in the context of old inflation \cite{SG}.

\begin{figure}
\begin{center}
\vspace{-10mm}
\epsfile{file=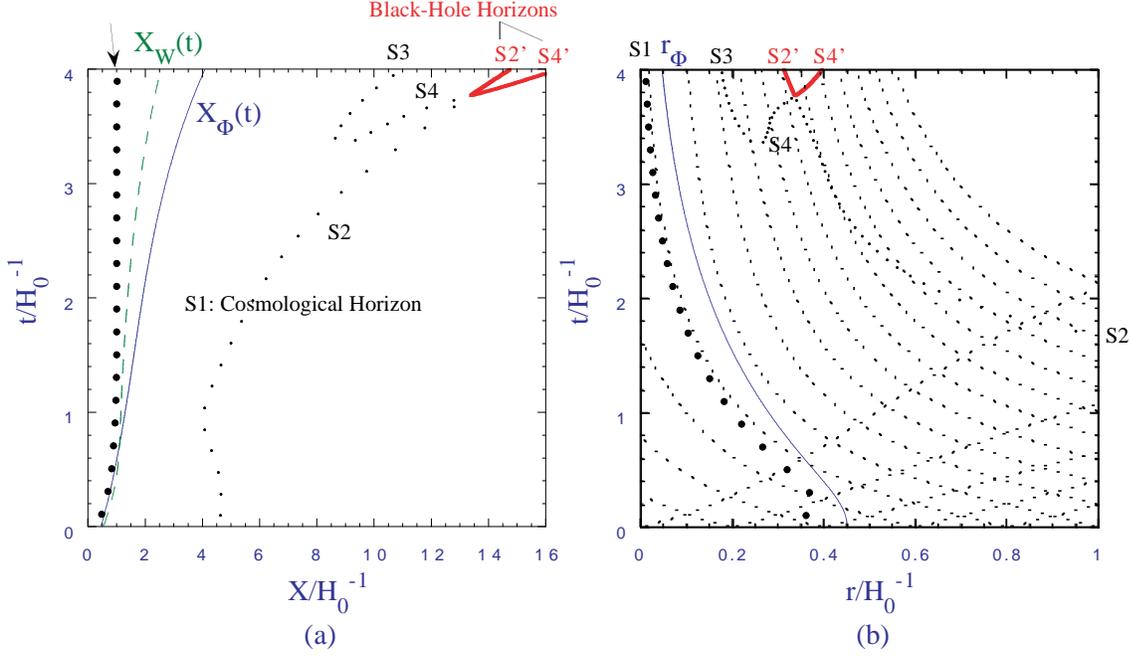,width=150mm}
\end{center}
\vspace{-5mm}
\caption{A solution of an inflating magnetic monopole.
(a) is the reproduction of Fig.\ 4(a) of \cite{Sak}:
we plot the trajectories of monopole boundaries and of apparent
horizons in terms of the proper distance from the center, $X$;
dotted lines denote apparent horizons;
we normalize time and length by the horizon scale defined as
$H_c^{-1}\equiv[8\pi V(0)/3\mpl^2]^{-1/2}$.
In (b) we plot ingoing and outgoing null geodesics (dotted lines) besides the
above-mentioned trajectories in a $t-r$ diagram; the boundary moves {\it
inward} and eventually becomes {\it spacelike}.}
\label{fig1}
\end{figure}

To see the causal structure of the inflating monopole more
closely, we also plot ingoing and outgoing null geodesics in Fig.\ 1(b).
An important feature for the magnetic monopole as well as the global
monopole \cite{CV} is that the boundary moves {\it inward}
and, moreover, it eventually becomes {\it spacelike}.

To understand the global structure, which is not completely covered by the
numerical solution, we make a reasonable assumption that the
core region approaches to de Sitter spacetime and the outside to
Reissner-Nordstr\"{o}m, as in Refs.\cite{TMT,BTV}.
In each static spacetime, apparent horizons and event horizons are
identical; the signs of $(\Theta^+,\Theta^-)$ are determined
as is shown in Fig.\ 2. The above assumption implies 
\begin{figure}
\begin{center}
\vspace{-4mm}
\epsfile{file=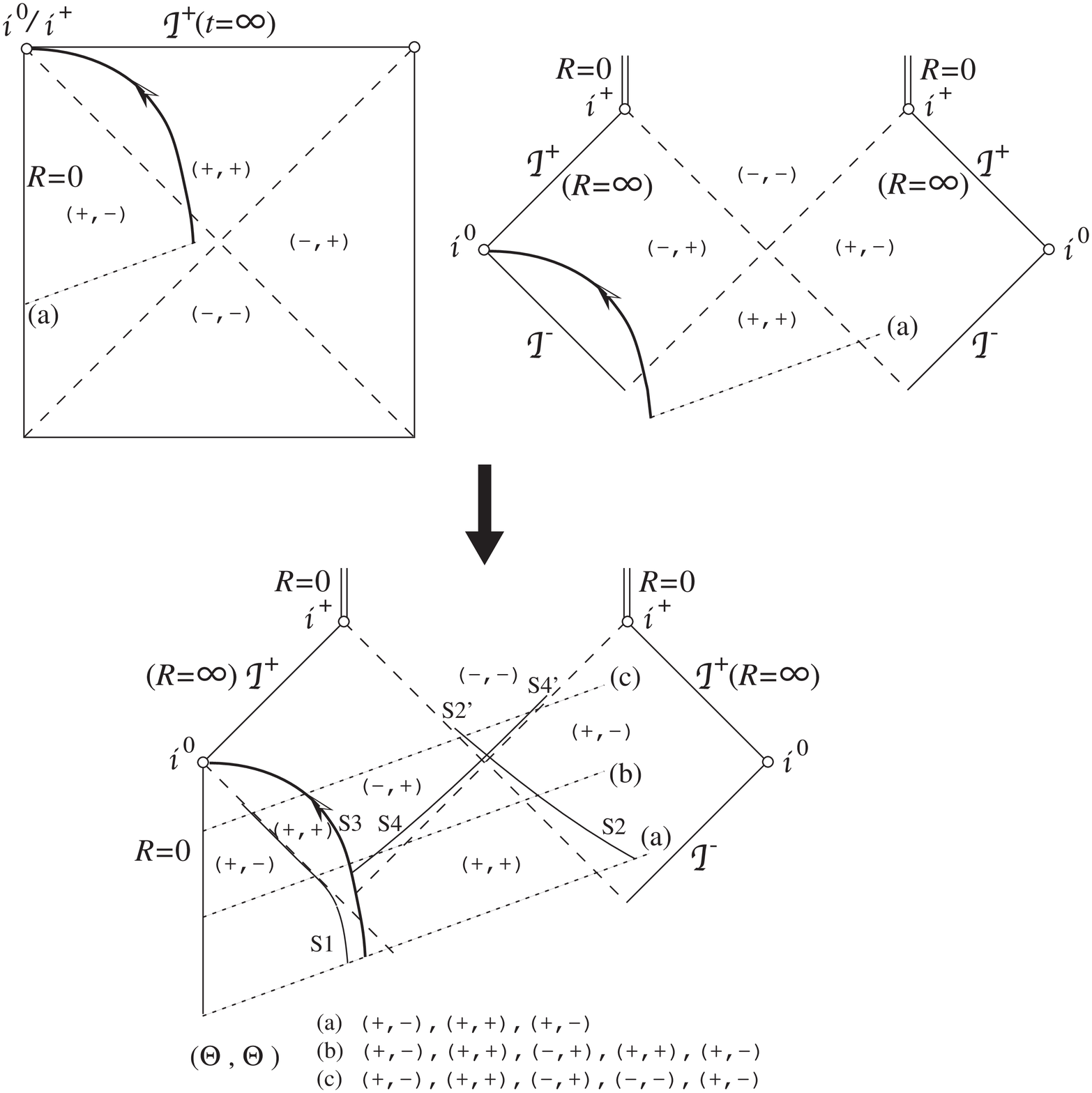,width=12cm}
\end{center}
\vspace{-3mm}
\caption{Possible conformal diagram for the spacetime of an inflating
magnetic monopole. The core region is approximated to be de Sitter spacetime
and the outside to be Reissner-Nordstr\"{o}m.
The upper figure shows how the monopole boundary is embedded in each
spacetime separately, and the lower figure shows a complete spacetime.
${\cal I}^+$ and ${\cal I}^-$ represent future and past null
infinity, $i^+$ represents future timelike infinity, and $i^0$
represents spacelike infinity.
Long-dashed lines denote event horizons, which are identified with
apparent horizons in de Sitter or Reissner-Nordstr\"{o}m spacetime.
Short-dashed lines (a), (b), and (c) denote time-slices, which
correspond to the embedding diagrams (a), (b), and (c) of Fig. 5 of 
\cite{Sak}, respectively.
For reference, we schematically depict the trajectories of apparent horizons.
We also write down the signs of the expansion $\Theta$.
The structure of apparent horizons, which is determined by the signs of
$\Theta^{\pm}$, completely agrees with that in Fig.\ 1 and Fig. 5 of 
\cite{Sak}.}\label{fig2}

\vskip 5mm
that, although the whole space in the
early stage is quite dynamical and an apparent horizon does not
coincide with an event horizon, the two horizons later approach
each other. Hence, we can extrapolate
the global structure from the structure of apparent horizons in the local
numerical solution.
From the consistency of the spatial distribution of the signs of
$(\Theta^+,\Theta^-)$, we conclude that Fig.\ 2 gives the only possible
embedding.

~~~~~Cosmologically, Fig.\ 2 tells us that any inflationary region eventually
enters a reheating phase because any observer inside the core
finally goes out. On the other hand, the monopole boundary continues
to expand in terms of the physical size and approaches spatial
infinity ($i^0$).
Figure 2 thus proves that this model is free from the ``graceful exit''
problem,
while the monopole itself undergoes ``eternal inflation''.
\end{figure}


\section{Topological Defect Inflation in Brans-Dicke Theory}

The Brans-Dicke-Higgs system, which we consider here, is described by the
action
\begin{equation}\label{action}
  S=\int d^4 x \sqrt{-g} \left[\frac{\Phi}{16\pi}{\cal R}
     -\frac{\omega}{16\pi\Phi}(\nabla_{\mu}\Phi)^2
     -\frac12(\nabla_{\mu}\phi^a)^2-V(\phi)\right],
~~ {\rm with} ~~
V(\phi)= {\lambda\over 4}(\phi^a\phi^a-\eta^2)^2,
\end{equation}
where $\Phi$ and $\omega$ are the BD field and the BD parameter, respectively.

In extended inflation \cite{LS}, where the BD theory was originally
introduced in inflationary cosmology, the universe does not 
expand exponentially but expands with a power law.
This slower expansion solved the graceful exit problem of old
inflation \cite{SG}. We thus expect 
that the BD field also affects the dynamics and global spacetime structure 
of inflating monopoles. 

Let us begin with a discussion of the fate of an inflating topological 
defect.
Once inflation begins, the core region can be approximated to be a 
homogeneous spacetime, where the scale factor and the BD field are
described by the solution of extended inflation \cite{LS}: 
$a(t)\propto t^{\omega+\frac12}, \Phi(t)\propto t^2.$
Hence, the effective Planck mass,
\begin{equation}
  \mpl(\Phi)=\sqrt{\Phi}\propto a^{{1\over\omega+\frac12}}, 
\end{equation}
continues to increase until inflation ends. This implies that, even if 
$\eta/\mpl(\Phi)$ is large enough to start inflation initially, it eventually
becomes smaller than the critical value 
$(\approx 0.33)$. We thus speculate that any defect comes to shrink
after inflation. 

In order to ascertain the above argument, we carry out numerical analysis 
for spherical global monopoles, using the coordinate system (\ref{metric}).
The above arguments are verified by this numerical result: as  
$\eta/m_{\rm Pl}(\Phi)$ gets close to the critical value, the monopole
stops expanding and begins to shrink. 

Figure 3 shows the trajectories of the position of $\phi=\eta/2$ 
for several values of $\omega$. 
The curves indicate that a monopole does 
not shrink toward the origin but tends to take a stable configuration.

The result that inflation eventually ends stems merely from the change 
of the local values of $m_{\rm Pl}(\Phi)$. We may therefore extend this 
result to models of other topological defects.

\begin{figure}
\begin{center}
\vspace{-11mm}
\epsfile{file=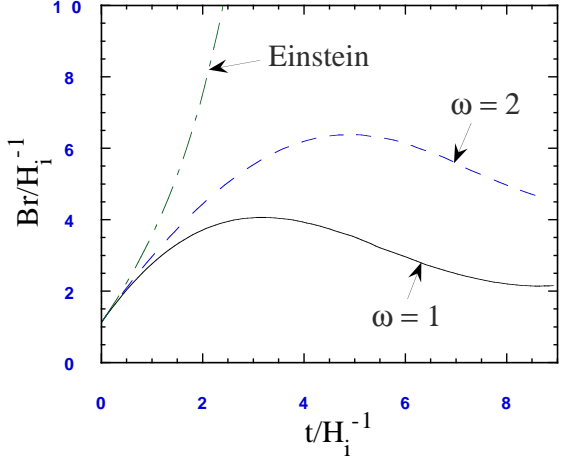,width=10cm}
\end{center}
\vspace{-4mm}
\caption{
Dynamics of a global monopole for several values of 
$\omega$. We set $\lambda=0.1$ and $\eta/m_{\rm Pl}(\Phi_i)=1$. We plot
trajectories of the position of $\phi=\eta/2$.
}\label{fig3}
\end{figure}

Next, we discuss constraints for this model to be a realistic
cosmological model.
One of the distinguishing features of this model is that there are two 
scenarios of exiting from an inflationary phase and entering a 
reheating phase. This is illustrated by the slow-roll conditions,
\begin{equation}
\epsilon^{-1}\equiv{\sqrt{6\pi}\eta\over m_{\rm Pl}(\Phi)}\gg 1,
~~ {\rm with} ~~
\phi\ll\eta\left(1-{\epsilon\over 2}\right)\equiv\phi_f.
\end{equation}
When the condition $\epsilon^{-1}\gg1$ (a more precise condition is 
$\eta/m_{\rm Pl}(\Phi)>0.33$) breaks down, inflation stops globally. Before
this 
time, many local regions with the present-horizon size enter a 
reheating phase when the second condition $\phi\ll\phi_f$ breaks down. The 
second scenario is just like that of standard TDI or other 
slow-roll inflationary models. In the first scenario a microscopic monopole 
might remain in the observable universe. Unfortunately, however, this 
possibility turns out to be ruled out because 
$\eta/m_{\rm Pl}(\Phi_f)$ must be larger than 1 from the COBE normalization, as
will be seen in Fig.\ 4(a) later. Thus we only consider the second
(standard) reheating scenario.

Starobinsky \& Yokoyama \cite{StY} derived the amplitude of a density
perturbation 
on comoving scale $l=2\pi/k$ in terms of Bardeen's variable $\Phi_A$ \cite{Bar} 
as
\begin{equation}\label{StY}
\Phi_A(l)^2={48 V\over 25m_{\rm Pl}(\Phi)^4}
\left[{8\pi V^2\over m_{\rm Pl}(\Phi)^2 V_{,\phi}^{~2}}
+{(e^{2\gamma^2 N}-1)^2\over 4\gamma^2}\right],
~~ {\rm with} ~~ 
\gamma\equiv\sqrt{{2\over2\omega+3}}
\end{equation}
where all quantities are defined at the 
time $t_k$ when the $k$-mode leaves the Hubble horizon, {\it i.e.}, when
$k=aH$. 
$N$ is defined as $N\equiv\ln(a_f/a_k)$, where the subscript $f$ 
denotes the time at the end of inflation.

Since the large-angular-scale anisotropy of the microwave background 
due to the Sachs-Wolfe effect is given by $\delta T/T=\Phi_A/3$, we 
can constrain the values of $\lambda$ and $\eta/m_{\rm Pl}$ by the 4yr
COBE-DMR data normalization \cite{Ben}:
$\Phi_A/3\cong 10^{-5}$, on the relevant scale.
We choose $N_{k=a_0H_0}=65$ typically, where a subscript 0 denotes 
the present epoch.
Figure\ 4(a) shows the allowed values of 
$\lambda$ and $\eta/m_{\rm Pl}$ for $\gamma=0.045$ ($\omega=500$). The 
concordant values are represented by two curves, which
is a distinguishing feature for the double-well potential. 
Unfortunately, fine-tuning of $\lambda\lesssim 10^{-13})$ is needed, just 
like other models \cite{StY}.
The constraints for $\omega>500$ are practically no 
different from those in the Einstein gravity. 

Using the relation $d\ln k=da/a+dH/H$ and the field equations, we 
obtain \cite{LL}
\begin{equation}\label{index}
n-1\equiv{d\ln\Phi_A^2\over d\ln k}
\cong-{3m_{\rm Pl}(\Phi)^2 V_{,\phi}^{~2}\over 8\pi V^2}
+{m_{\rm Pl}(\Phi)^2 V_{,\phi\phi}\over 4\pi V}
-6\gamma^2.
\end{equation}
The spectral indices are plotted in Fig.\ 4(b); the two lines correspond 
to the two lines of COBE-normalized amplitudes in Fig.\ 4(b). As Eq.\
(\ref{index}) suggests, the deviation from the Einstein theory is only 
$6\gamma^2\cong 0.01$ for $\omega=500$. Therefore, the relatively large shift 
from $n=1$ is caused not by the BD field but by the double well potential.

Finally, we consider the detectability of relic monopoles in this 
model. Due to the sufficient expansion, $N>65$, the core of a monopole 
is located so far from the observed region that even the long-range term 
of a global monopole, $\rho=\eta^2/R^2$, does not exert any 
effective astrophysical influence.

\begin{figure}
\begin{center}
\vspace*{-5mm}
\epsfile{file=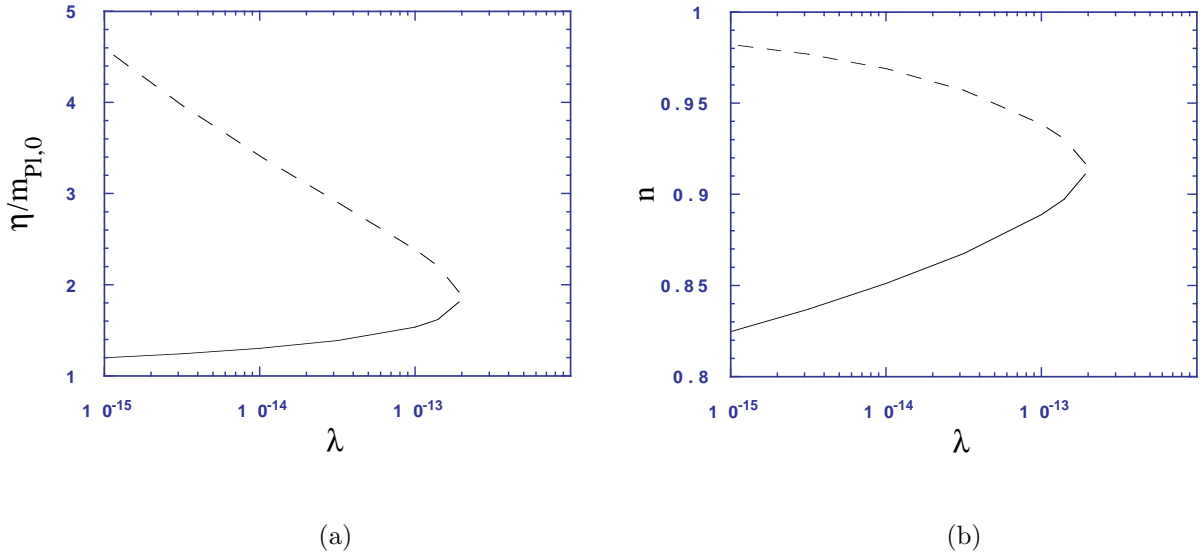,width=16cm}
\end{center}
\vspace{-1mm}
\hspace*{43mm}(a)\hspace{79mm}(b)
\caption{
Constraints from COBE-normalized amplitudes of density perturbations.
(a) Concordant values of $\lambda$ and $\eta/m_{\rm Pl,0}$.
(b) The spectral indices of density perturbations.
}\label{fig4}
\end{figure}

\section{Topological Defect Inflation Induced by Non-Minimally Coupled 
Massive Scalar Field}

The model with a non-minimally coupled massive scalar field is 
described by the action,
\begin{equation}\label{massive}
{\cal S} =\int d^4 x \sqrt{-g} \left[\frac{\cal R}{2\kappa^2}
    -\frac12\xi\phi^2{\cal R}-\frac12(\nabla\phi)^2-V(\phi)\right],
~~ {\rm with} ~~ V(\phi)=\frac12 m^2\phi^2
~~ {\rm and} ~~ \kappa^2\equiv{8\pi\over{\mpl}^2}.
\end{equation}
Futamase \& Maeda \cite{FM} investigated how the nonminimal coupling 
term $(1/2)\xi{\cal R}\phi^2$ affects realization of chaotic inflation
\cite{Lin}, and derived constraints on the coupling 
constant $\xi$ from the condition for sufficient inflation. For the 
massive scalar model (\ref{massive}), they obtained $|\xi|\lesssim 10^{-3}$. 
Here we  show that another type of inflation may be possible if 
$\xi<0$, which relaxes the constraint on $\xi$. 

\begin{figure}
\begin{center}
\vspace*{-9mm}
\epsfile{file=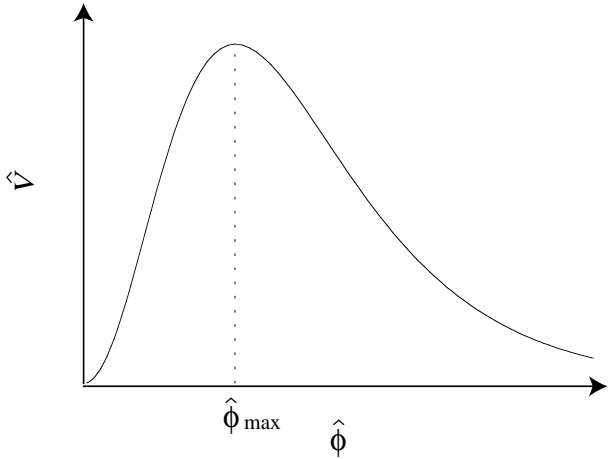,width=80mm}
\end{center}
\vspace{-4mm}
\caption{
The potential $\hat V(\hat\phi)$ in a fictitious world for
$\xi<0$ \cite{FM}. A plateau at $\hat\phi=\hat\phi_{\rm max}$
leads TDI.}\label{fig5}
\end{figure}

Following Futamase \& Maeda, we apply the conformal transformation
and redefine a scalar field so that the model is described by the 
Einstein gravity with a canonical 
scalar field:
\begin{equation}\label{hatS}
  \hat{\cal S} = \int d^4 \hat x \sqrt{-\hat g}
    \biggl[\frac{\hat{\cal R}}{2\kappa^2}-{1\over2}(\hat\nabla\hat\phi)^2
     -\hat V(\hat\phi) \biggr],
~~ {\rm with} ~~
\hat V(\hat\phi)={V(\phi)\over(1+\psi)^2}
~~ {\rm and} ~~ \psi\equiv-\xi\kappa^2\phi^2.
\end{equation}
Thanks to this standard form of the theory, we can discuss 
the qualitative behavior of $\hat\phi$ (or $\phi$) in terms of the potential 
shape. We depict $\hat V(\hat\phi)$ for $\xi<0$ in Fig.\ \ref{fig5}. 
A distinguishing feature of this potential is that it has a maximum 
at $\hat\phi=\hat\phi_{\rm max}$ corresponding to $\psi=1$ or
$\phi=1/\kappa\sqrt{-\xi}\equiv\phi_{\rm max} $. Hence, if the initial 
value of the scalar field, $\hat\phi_i$, is larger than $\hat\phi_{\rm max}$
and 
if the energy density of the scalar field
$E_{\hat\phi}=\hat\phi_{,\hat t}^{~2}/2+\hat V(\hat\phi)$ is below 
$\hat V(\hat\phi_{\rm max})$, $\Phi$ cannot reach the origin
$\hat\phi=\phi=0$, but it will run away to 
infinity as long as the universe is expanding in the conformal frame 
with $\hat H>0$. This runaway solution is irrelevant to our Universe, 
and Futamase \& Maeda \cite{FM} obtained a constraint $|\xi|<10^{-3}$ 
so that $\hat\phi_{\rm max}>5\mpl$ and the initial value of $\hat\phi$ 
required for sufficient chaotic inflation $\hat\phi_i\sim 5\mpl$ lies on 
the left of the potential peak.

We can argue, however, that sufficient inflation may be possible even if 
$\hat\phi_{\rm max}<5\mpl$ because the plateau around 
$\hat\phi=\hat\phi_{\rm max}$ may
cause another channel of inflation. Seen in the conformal frame, some 
domains of the universe relax to $\hat\phi=0$ and others run away to 
$\hat\phi\rightarrow\infty$ as the universe expands. In between these two 
classes of regions exist domain walls where large potential energy 
density $\hat V(\hat\phi_{\rm max})$ is stored. If the curvature of 
the potential is sufficiently small there, such domain walls will 
inflate. This is nothing but TDI.

Let us discuss whether inflation can take place at 
$\hat\phi=\hat\phi_{\rm max}$,
following the arguments of Linde \& Vilenkin \cite{LV}.
The standard conditions for slow-roll inflation lead to
\begin{equation}\label{cond1}
{|\xi|\over1+3|\xi|}\ll\frac 34.
\end{equation}
A similar relation is derived from the condition that the thickness of the 
wall characterized by the curvature scale of the potential at the 
maximum, $\hat R_w$, is greater than the horizon, $\hat H^{-1}$:
\begin{equation}\label{cond2}
\hat R_w\hat H=
\sqrt{\kappa^2\hat V(\hat\phi_{\rm max})\over 3\hat
V_{,\hat\phi\hat\phi}(\phi_{\rm max})}
=\sqrt{{1+3|\xi|\over 12|\xi|}}\gtrsim 1.
\end{equation}
Inequalities (\ref{cond1}) and (\ref{cond2}) suggest that inflation actually 
takes place at the top of the potential if $|\xi|\ll 1$. Once inflation 
sets in, it continues forever inside a domain wall. 

In order to obtain more precise conditions for sufficient inflation, we 
solve the (homogeneous and isotropic) field equations numerically. We assume
initial 
values as $\hat\phi_i=\delta\hat\phi_{\rm Q}\equiv\hat H/2\pi$ and 
$(\hat\phi_{,\hat t})_i=0$, and observe the e-fold number of inflation after the
classical 
dynamics dominates over quantum fluctuations, {\it i.e.}, 
$|\hat\phi_{,\hat t}|/\hat H>\delta\hat\phi_{\rm Q}$. Sufficient expansion
typically requires ${a_f/a_c}>e^{65}$, 
where $c$ denotes the time when $|\hat\phi_{,\hat t}|/\hat H=\delta\hat\phi_{\rm
Q}$.
The allowed region is plotted in Fig.\ 6(a). This shows that inflation
continues 
sufficiently even if $|\xi|\cong0.1$, contrary to the previous result
\cite{FM}.

Next, we investigate density perturbations generated during 
inflation, and constrain the model from the 4yr COBE-DMR data 
\cite{Ben}. Because Makino \& Sasaki \cite{MS} showed that the density 
perturbation in the original frame exactly coincides with that in the 
conformal frame, we can easily calculate the amplitude and the 
spectral index with the well-known formulas. The amplitude of 
perturbation is given as
\begin{equation}
\Phi_A\left(l={2\pi\over k}\right)
={\sqrt{3}\kappa^3\hat V^{\frac32}\over10\pi\hat V_{,\hat\phi}}=
{\sqrt{3}\kappa^3m\hat\phi^2\sqrt{1+(\xx)\psi}\over20\sqrt{2}\pi(1-\psi^2)}.
\end{equation}
The large-scale anisotropy of the cosmic 
microwave background measured by COBE-DMR leads to \cite{Ben}
$\delta T/T=\Phi_A/3\cong 10^{-5}$.
The concordant values of $\xi$ and $m$ are also plotted in Fig.\ 6(a). 

\begin{figure}
\begin{center}
\vspace{-3mm}
\epsfile{file=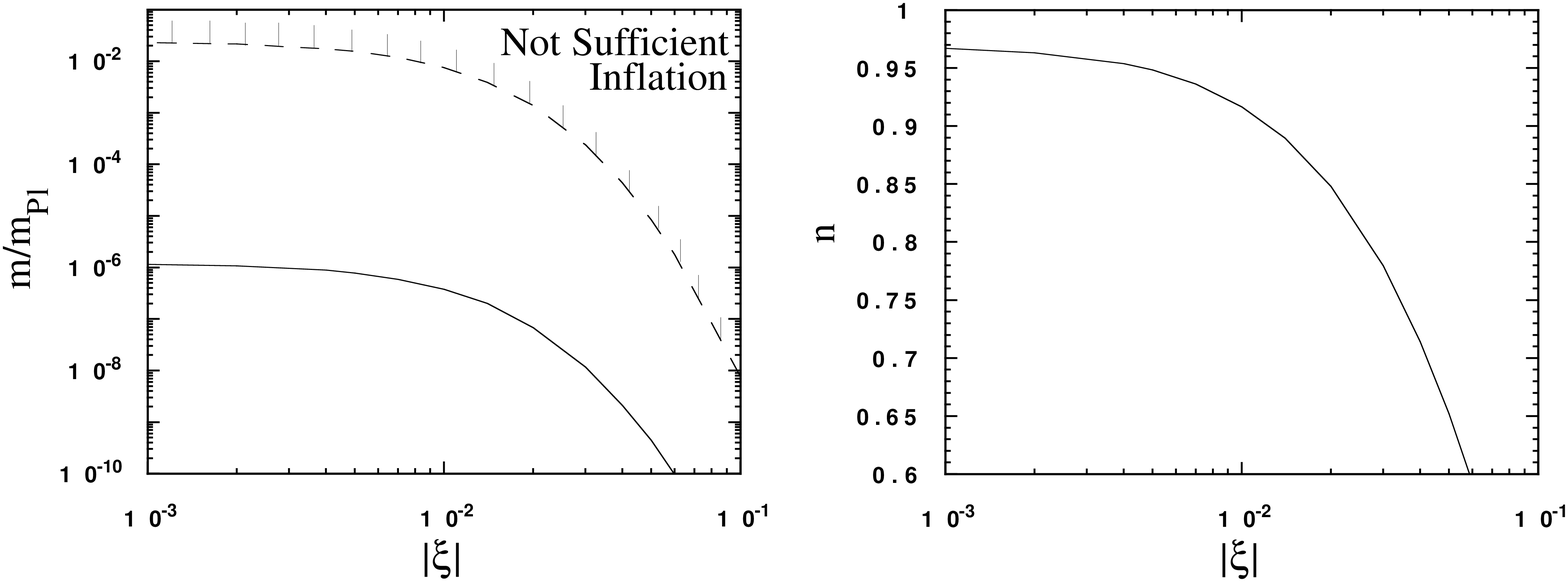,width=160mm}
\end{center}
\vspace{-1mm}
\hspace*{43mm}(a)\hspace{77mm}(b)
\caption{
Constraints from COBE-normalized amplitudes of density perturbations.
(a) Constraints on $\xi$ and $m/\mpl$. The 
dashed curve is a constraint from sufficient inflation. The 
solid curve represents the concordant values with the 
amplitude of the COBE-DMR data.
(b) The spectral indices of density perturbations.
}\label{fig6}
\end{figure}

The spectral index $n$ is given as \cite{LL}
\begin{equation}
n-1\equiv{d\ln \Phi_A^2\over d\ln k}
={4\over\kappa^2\hat\phi^2}\left[-{3(1-\psi)^2\over 1+(\xx)\psi}
+{1-6\psi-(5-35\xi)\psi^2+2(\xx)\psi^3\over\{1+(\xx)\psi\}^2}\right].
\end{equation}
The values of $n$ which satisfy the COBE-DMR normalization are plotted 
in Fig.\ \ref{fig6}(b). With the observational data,
$n=1.2\pm0.3$ \cite{Ben}, we have a constraint $|\xi|\lesssim10^{-2}$, which 
is still less stringent than the previous result \cite{FM}.

\section*{Acknowledgments}

I would like to thank Tomohiro Harada, Kei-ichi Maeda, Ken-ichi Nakao, 
and Jun'ichi Yokoyama, with whom the present work was done.
Thanks are also due to Paul Haines for reading the manuscript.
Numerical Computation of this work was carried out at the Yukawa Institute
Computer Facility. 
I was supported by JSPS Research Fellowships for Young Scientist, No.\ 9702603.



\begin{thebibliography}{99}
\bibitem{GR}E.I. Guendelman \& A. Rabinowitz, Phys. Rev. D. {\bf 44}, 3152
(1991).
\bibitem{LV}A. Linde, Phys. Lett. B {\bf 327}, 208 (1994);
A. Vilenkin, Phys. Rev. Lett. {\bf72}, 3137 (1994).
\bibitem{SSTM}N. Sakai, H. Shinkai, T. Tachizawa \& K. Maeda, Phys.
Rev. D {\bf 53}, 655 (1996).
\bibitem{Sak}N. Sakai, Phys. Rev. D {\bf 54} 1548 (1996).
\bibitem{EI}J. Ellis, N. Kaloper, K.A. Olive \& J. Yokoyama, Phys.
Rev. D {\bf 59}, 103503 (1999);\\
Izawa K.-I., M. Kawasaki \& T. Yanagida, Prog. Theor. Phys. {\bf101}, 1129
(1999).
\bibitem{SNH}N. Sakai, K. Nakao \& T. Harada, preprint gr-qc/9909027.
\bibitem{SYM}N. Sakai, J. Yokoyama \& K. Maeda, Phys. Rev. D {\bf 59}, 
103504 (1999).
\bibitem{SY}N. Sakai \& J. Yokoyama, Phys. Lett. B {\bf 456}, 113 
(1999).
\bibitem{VG}For a review of spacetime solutions with non-Abelian fields, see,
e.g.,
M.S. Volkov \& D.V. Gal'tsov, preprint hep-th/9810070.
\bibitem{OLB}M.E. Ortiz, Phys. Rev. D {\bf 45}, R2586 (1992);\\
K. Lee, V.P. Nair \& E.J. Weinberg, {\it ibid.} {\bf 45}, 2751 (1992);\\
P. Breitenlohner, P. Forg$\grave{{\rm a}}$cs \& D. Maison, Nucl. Phys.
{\bf B383}, 357 (1992); {\it ibid.} {\bf B442}, 126 (1995).
\bibitem{TMT}T. Tachizawa, K. Maeda \& T. Torii, Phys. Rev. D {\bf 51},
4054 (1995).
\bibitem{BTV}A. Borde, M. Trodden \& T. Vachaspati, Phys. Rev. D
{\bf 59}, 43513 (1998);\\
T. Vachaspati \& M. Trodden, preprint gr-qc/9811037.
\bibitem{NOK}T. Nakamura, K. Oohara \& Y. Kojima, Prog. Theor. Phys. Suppl.
{\bf 90}, 1 (1987).
\bibitem{SSKM} K. Sato, M. Sasaki, H. Kodama \& K. Maeda, Prog. Theor. Phys.
{\bf 65}, 1443 (1981);\\
K. Sato, H. Kodama, M. Sasaki \& K. Maeda, {\it ibid.} {\bf 108B}, 
103 (1982).
\bibitem{SG}K. Sato, Mon. Not. Roy. Astron. Soc. {\bf 195}, 467 (1981);
Phys. Lett. {\bf 99B}, 66 (1981);\\
A.H. Guth, Phys. Rev. D {\bf 23}, 347 (1981).
\bibitem{CV}I. Cho \& A. Vilenkin, Phys. Rev. D {\bf 56}, 7621 (1997).
\bibitem{LS}D. La \& P.J. Steinhardt, Phys. Rev. Lett. {\bf 62}, 376 (1989).
\bibitem{StY}A. Starobinsky \& J. Yokoyama, Proc. 4th Workshop on {\it
General Relativity and Gravitation} eds. K. Nakao {\it et al.} (Yukawa
Institute for Theoretical Physics), 381 (1994).
\bibitem{Bar}J.M. Bardeen, Phys. Rev. D {\bf 22}, 1882 (1980).
\bibitem{Ben}C.L. Bennet {\it et al.}, Astrophys. J. {\bf 464}, L1 (1996).
\bibitem{LL}A.R. Liddle \& D.H. Lyth, Phys. Lett. B {\bf 291}, 391 (1992).
\bibitem{FM}T. Futamase \& K. Maeda, Phys. Rev. D {\bf 39}, 399 (1989).
\bibitem{Lin}A. Linde, Phys. Lett. B {\bf 129}, 177 (1983).
\bibitem{MS}N. Makino \& M. Sasaki, Prog. Theor. Phys. {\bf 86}, 103 (1991).
\end{thebibliography}
\end{document}